\documentclass[onecolumn,showpacs,preprintnumbers,amsmath,amssymb,aps]{revtex4}
 \normalsize

\usepackage{graphicx,graphics}
\usepackage{dcolumn}
\usepackage{bm}
\usepackage{amssymb}
\usepackage{amsmath}
\newcommand{\bi}{\bibitem}
\newcommand{\be}{\begin{eqnarray}}
\newcommand{\ee}{\end{eqnarray}}
\newcommand{\nn}{\nonumber}

\def\hbar#1{\backslash\hspace{-2mm}#1}

\newcommand{\lsim}{\raise0.3ex\hbox{$\;<$\kern-0.75em\raise-1.1ex\hbox{$\sim\;$}}}
\def\nn{\nonumber}

\def\2tvec#1#2{
\left(
\begin{array}{c}
#1  \\
#2  \\
\end{array}
\right)}

\def\mat2#1#2#3#4{
\left(
\begin{array}{cc}
#1 & #2 \\
#3 & #4 \\
\end{array}
\right) }

\def\Mat3#1#2#3#4#5#6#7#8#9{
\left(
\begin{array}{ccc}
#1 & #2 & #3 \\
#4 & #5 & #6 \\
#7 & #8 & #9 \\
\end{array}
\right) }

\def\3tvec#1#2#3{
\left(
\begin{array}{c}
#1  \\
#2  \\
#3  \\
\end{array}
\right)}

\def\hbar#1{\backslash\hspace{-2mm}#1}

\def\bea{\begin{eqnarray}}
\def\eea{\end{eqnarray}}

\def\nn{\nonumber}

\newcommand{\bers}{\begin{eqnarray*}}
\newcommand{\eers}{\end{eqnarray*}}
\newcommand{\bt}{\begin{itemize}}
\newcommand{\et}{\end{itemize}}

\begin{document}
\preprint{KANAZAWA-11-02} \vfill \preprint{} \vspace{2.0cm}
\title{Right-handed Sneutrino Dark Matter in Supersymmetric $B-L$ Model}

\author{Shaaban Khalil}
\affiliation{Centre for Theoretical Physics, The British University in Egypt,
El Sherouk City, Postal No.~11837, P.O.~Box 43, Egypt}
\affiliation{Department of Mathematics, Ain Shams University,
Faculty of Science, Cairo 11566, Egypt}

\author{Hiroshi Okada}
\affiliation{Centre for Theoretical Physics, The British University in Egypt,
El Sherouk City, Postal No.~11837, P.O.~Box 43, Egypt}

\author{Takashi Toma}
\affiliation{
 Institute for Theoretical Physics, Kanazawa University, Kanazawa
  920-1192, Japan}

\begin{abstract}
We show that the lightest right-handed sneutrino in {\rm TeV}
scale supersymmetric $B-L$ model with inverse seesaw mechanism is
a viable candidate for cold dark matter. We find that it accounts
for the observed dark matter relic abundance in a wide range of
parameter space. The spin-independent cross section of $B-L$
right-handed sneutrino is consistent with the recent results CDMS
II and XENON experiments and it is detectable in future direct
detection experiments. Although the $B-L$ right-handed sneutrinos
annihilate into leptons, the PAMELA results can not be explained
in this model unless a huge boost factor is considered. Also the
muon flux generated by $B-L$ right-handed sneutrino in the
galactic center is smaller than Super-Kamiokande's upper bound.
\end{abstract}

\pacs{ }

\maketitle
\section{Introduction}

The experimental verifications of non-vanishing neutrino masses
and the alluring hints of dark matter's (DM's) existence are
serious indications for new physics beyond the Standard Model
(SM). Supersymmetry (SUSY) is an attractive candidate for new
physics at {\rm TeV} scale that provides an elegent solution for
the SM gauge hierarchy problem and stabilize the SM Higgs mass at
the electroweak scale. The minimal supersymmetric standard model
(MSSM) is the simplest extension of the SM. In order to account
for the observed neutrino masses and mixing, SM singlets
(right-handed neutrinos) are usually introduced.

In the MSSM with $R$-parity conservation and universal soft SUSY
breaking terms, the lightest neutralino, which is typically bino
dominated, is an attractive candidate for cold DM. However, the
current experimental constraints on SUSY particles lead to
overproduction of bino relic abundance, in contradiction with the
observational limits of the Wilkinson Microwave Anisotropy Probe
(WMAP) \cite{Spergel:2006hy}. In addition, the recent results of
Cryogenic DM Search (CDMS II) \cite{Ahmed:2009zw} set an upper
limit on the DM-nucleon elastic scattering spin independent cross
section of order $3.8 \times 10^{-44} {\rm cm}^2$ for DM mass of
$70$ {\rm GeV}, which imposes stringent limits on the lightest
neutralino of the MSSM, even if it consists of gaugino-Higgsino
mixture. Therefore, one concludes that the DM constraints severely
reduce the allowed range in the parameter space of the MSSM. It is
also worth mentioning that the observed anomalies in the cosmic
rays may favor the type of dark matter that annihilates into
leptons not to quarks, unlike the lightest neutrlino in MSSM.

{\rm TeV} scale right-handed neutrinos can be naturally implemented in
supersymmeric $B-L$ extension of the SM (SUSY $B-L$), which is
based on the gauge group $G_{B-L}\equiv SU(3)_C \times SU(2)_L
\times U(1)_Y \times U(1)_{B-L}$ \cite{Khalil:2006yi}. In this
model, three SM singlet fermions arise quite naturally due to the
$U(1)_{B-L}$ anomaly cancellation conditions. These particles are
accounted for right-handed neutrinos, and hence a natural
explanation for the seesaw mechanism is obtained
\cite{Khalil:2006yi,Abbas:2007ag,Emam:2007dy}. 
This work was based on earlier papers in Ref. \cite{b-l-origin}.
In this class of
model, the scale of $B-L$ symmetry breaking is related to
supersymmetry breaking scale \cite{Khalil:2007dr}. Therefore, the
right-handed neutrino masses are naturally of order {\rm TeV} scale.
This has initiated a considerable interest in analyzing the
phenomenological implications of these models and their possible
signatures at the LHC \cite{otherb-l}.

{\rm TeV} scale SUSY $B-L$ is one of the simplest models that
provides viable and testable solution to the two puzzles of the DM
and the neutrino masses. However, in order to fulfill the
experimental measurements for the light-neutrino masses, one of
the following scenarios must be adopted: $(i)$ Type I seesaw
mechanism with very small Dirac neutrino Yukawa couplings,
$Y_{\nu} < {\cal O}(10^{-7})$ \cite{Khalil:2006yi}. $(ii)$ Inverse
seesaw mechanism with order one Yukawa couplings and small mass
scale $\sim {\cal O}(1)$ {\rm KeV}, that corresponds to scale of
breaking a remnant discrete symmetry, $(-1)^{L+S}$
\cite{Khalil:2010iu}. This work was based on earlier papers in Ref. \cite{inverse-origin}.
In the first case, due to the smallness of
Dirac Yukawa couplings, the right-handed neutrino sector has a
very suppressed interaction with the SM particle. Therefore, the
prediction of SUSY $B-L$ remains close to the MSSM ones. It turns
out that the DM candidate of this model is still the lightest
neutralino \cite{Khalil:2008ps}, which is a kind of a mixture of
three neutral gauginos and four neutral Higgsino.

In this paper we consider the scenario where the right-handed
sneutrino in SUSY $B-L$ with inverse seesaw is the lightest SUSY
particle (LSP) and stable, so that it can be a cold DM candidate \cite{ArkaniHamed:2000kj}.
It is worth mentioning that in MSSM the left-handed sneutrino is
the only stable weakly coupled neutral boson that can be a DM
candidate \cite{Hall:1997ah}. However, the current limits in direct detection
experiments rule out this possibility, since left-handed sneutrino
has a tree level interaction with the $Z$ gauge boson, hence its
elastic scattering cross section with nucleons is quite large. In
case of MSSM extension with {\rm TeV} right-handed neutrino ($N$)
superfields, the interaction of $N$ with the SM particles can be obtained from the superpotential:%
\be %
{\cal W}= {\cal W}_{\rm MSSM} +  Y_{\nu} N^c L H_2. %
\label{mssm} %
\ee %
Due to smallness of $Y_\nu$, the annihilation cross section of the
right-handed sneutrinos is very suppressed, hence its relic
density is larger than the WMAP limit. Moreover, since they cannot
couple to the quark sectors, the direct detection experiments such
as CDMS II cannot be tested in this case.

We will show that in the SUSY $B-L$ model with $Y_\nu \sim {\cal
O}(1)$ the dominant annihilation channel for right-handed
sneutrino is given by the four point interactions, leading to
$h^0$ and $h^0$. While the effective couplings of right-handed
sneutrinos with quarks are obtained through the exchange of {\rm
TeV} scale $B-L$ gauge boson, $Z_{B-L}$. Thus, one finds that the
WMAP results of the DM relic abundance and CDMS II/XENON for
direct detection can be accommodated in SUSY $B-L$ simultaneously
\cite{bl-dm}.

The paper is organized as follows. In section 2 we analyze the
supersymmetric $B-L$ model with Inverse Seesaw Mechanism. We study
the neutrino sector and the neutrino mass eigenstates. In section
3 we analyze the $B-L$ right-handed sneutrino mass and
interactions. Section 4 is devoted for the analysis of the relic
abundance of $B-L$ right-handed sneutrino. We show that the WMAP
limits can be easily satisfied in a wide range of parameter space.
In section 5 we discuss the direct detection rate of $B-L$
right-sneutrino DM. We show that the elastic cross section of our
DM candidate with nucleon is consistent with the recent results of
CDMS II/XENON experiment and it is detectable in near future
experiments. In section 6 we discuss the indirect detection rate
of $B-L$ right-handed sneutrino. We show that the annihilation
channels of right-handed sneutrino into leptons are subdominant,
therefore it cannot account for the controversial PAMELA results.
We also show that the muon flux generated from the right-handed
sneutrino in the galactic center is much smaller than the
Super-Kamiokande's limits. Finally we give our conclusions in
section 6.


\section{Supersymmetric $B-L$ model with Inverse Seesaw Mechanism}

The proposed {\rm TeV} scale supersymmetric $B-L$ extension of the SM is
based on the gauge group $SU(3)_C\times SU(2)_L\times U(1)_Y\times
U(1)_{B-L}$, where the $U(1)_{B-L}$ is spontaneously broken by a
chiral singlet superfield $\chi_1$ with $B-L$ charge $=+1$
and $\chi_2$ with $B-L$ charge $=-1$. As in the conventional $B-L$
model, a gauge boson $Z_{B-L}$ and three chiral singlet
sueperfields $N_{i}$ with $B-L$ charge $=-1$ are introduced for
the consistency of the model. Finally, three chiral singlet
superfields $S_1$ with $B-L$ charge $=+2$ and three chiral singlet
superfields $S_2$ with $B-L$ charge $=-2$ are considered to
implement the inverse seesaw mechanism.
The superpotential of the leptonic sector of this model is given by%
\be%
{\cal W}= Y_{e} E^c L H_1 + Y_{\nu}N^cLH_2  + Y_{S} N^c \chi_1
S_{2} + \mu H_1H_2 + \mu' \chi_1\chi_2.
\label{sp}
\ee
It is worth noting that the chiral singlet superfields $\chi_2$ and $N$ have the same $B-L$ charge. Therefore, one may impose a discrete symmetry in order to distinguish them and to prohibit other terms beyond those given in Eq. (\ref{sp}).
In this case, the relevant soft SUSY breaking terms, assuming the usual
universality assumptions, are as follows
 \bea - {\cal L}_{soft} &=&
\sum_{\phi} {\widetilde{m}}_{\phi}^{2} \vert \phi \vert^2 +
Y_{\nu}^{A}{\widetilde{N}}^{c} {\widetilde{L}}H_{2} +
Y_{e}^{A}{\widetilde{E}}^{c}{\widetilde{L}}H_{1}+Y_{S}^{A}{\widetilde{N}}^{c}{\widetilde
S_{2}} \chi_1  +B \mu H_1 H_2 + B\mu' \chi_1 \chi_2
 \nn\\&+& \frac{1}{2} M_1 {\widetilde{B}}{\widetilde{B}}+
 \frac{1}{2}M_2{\widetilde{W}}^a {\widetilde{W}}^a
+ \frac{1}{2}M_3 {\widetilde{g}}^a {\widetilde{g}}^a + \frac{1}{2}
M_{B-L} {\widetilde{Z}_{B-L}}{\widetilde{Z}_{B-L}}+ h.c ,
\label{soft}
\eea%
where the sum in the first term runs over $\phi=H_1, H_2, \chi_1,
\chi_2, \tilde L, \tilde E^c,\tilde N^c,\tilde S_1, \tilde S_2$
and $Y_L^A\equiv Y_L A_L$ ($L=e,\nu,S$) is the trilinear associated with lepton
Yukawa coupling.  In order to prohibit a possible large mass term
$M S_1 S_2$ in the above, we assume that the particles, $N^c_i$,
$\chi_{1,2}$, and $S_2$ are even under matter parity, while $S_1$
is an odd particle. The $B-L$ symmetry is radiatively broken by
the non-vanishing vacuume expectation values (VEVs) $\langle
\chi_1 \rangle = v^\prime_1$ and $\langle \chi_2 \rangle = v^\prime_2$
\cite{Khalil:2007dr}. The tree level
potential $V(\chi_1,\chi_2)$ is given by %
\be %
V(\chi_1,\chi_2)= \frac{1}{2} g^2_{B-L}(\vert
\chi_2\vert^2 - \vert \chi_1\vert^2)^2 + \mu^2_1 \vert
\chi_1\vert^2 + \mu^2_2\vert \chi_2\vert^2 - \mu^2_3(\chi_1 \chi_2
+
h.c).%
\ee %
At GUT scale, $\mu^2_i = m_0^2 + \mu^{\prime^2}, i=1,2$ and $\mu^2_3 = - B
\mu^\prime$.
However, they have different evolution from GUT scale to {\rm TeV} scale and $\mu_2$ becomes negative, so that $B-L$ is spontaneously broken \cite{Khalil:2007dr}.
The minimization of
$V(\chi_1,\chi_2)$ leads to the following condition:%
\be %
v^{\prime^2} = (v^{\prime^2}_1 + v^{\prime^2}_2) =
\frac{(\mu^2_1-\mu^2_2) - (\mu^2_1+\mu^2_2)\cos2\theta}{2
g^2_{B-L}
\cos2\theta},%
\label{muprime}
\ee %
The angle $\theta$ is defined as $\tan \theta =v'_1/v'_2$. The
minimization conditions also leads to
\be %
\sin 2 \theta = \frac{2
\mu^2_3}{\mu^2_1+\mu^2_2}.%
\ee %

After $B-L$ breaking, the $Z_{B-L}$ gauge boson acquires a mass
\cite{Khalil:2006yi}: $M^2_{Z_{B-L}}=4 g^2_{B-L}
v^{\prime^2}$. The high energy experimental searches for an extra
neutral gauge boson impose lower bounds on this mass. The
stringent constraint on $U(1)_{B-L}$ obtained from LEP II result,
which
implies \cite{m.carena} %
\be
\frac{M_{Z_{B-L}}}{g_{B-L}}>6\ {\rm TeV}.\label{z-b-l-constrain} %
\ee

Now we turn to the neutrino sector and show how the observed
light-neutrino masses can be obtained with ${\cal O}(1)$ Dirac
neutrino Yukawa coupling. As can be seen from Eq. (\ref{sp}), after
$B-L$ and EW symmetry breaking, the neutrino Yukawa interaction
terms lead to the following mass
terms:%
\be%
{\cal L}_m^{\nu} = m_D \bar{\nu}_L N^c + M_N N^c S_2,%
\ee%
where $m_D=Y_{\nu} v\sin\beta$ and $ M_N = Y_{S} v' \sin\theta$. From this Lagrangian, one can easily observe that
although the lepton number is broken through the spontaneous $B-L$
symmetry breaking, a remnant symmetry: $(-1)^{L+S}$ is survived,
where $L$ is the lepton number and $S$ is the spin. After this
global symmetry is broken at much lower scale, a mass term for
$S_2$ (and possibly for $S_1$ as well) is
generated. Therefore, the Lagrangian of neutrino masses, in the flavor basis, is given by: %
\be%
{\cal L}_m^{\nu} = m_D \bar{\nu}_L N^c   + M_N N ^c S_2 + \mu_{S_2} S_2^2(+ \mu_{S_1} S_1^2).%
\ee%
In the basis $\{\nu_L, N^c, S_2\}$, the $3\times 3$ neutrino mass
matrix of one generation takes the form:%
\be
\left(%
\begin{array}{ccc}
  0 & m_D & 0\\
  m_D & 0 & M_N \\
  0 & M_N & \mu_{S_2}\\
\end{array}%
\right). %
\ee%
The mixing matrix($O$) for this mass matrix leads to
the following light and heavy neutrino masses respectively in the
limit of $\mu_{S_i}<<m_D,~M_N$ (where $i=1,2$): \be
m_{\nu_{\ell}}=\frac{m_D^2\mu_{S_2}}{M_N^2+m_D^2},~ m_{\nu_{H,H'}}=\pm
\sqrt{M_N^2+m_D^2}+\frac{1}{2}\frac{M_N^2 \mu_{S_2}}{M_N^2+m_D^2}, \ee
 where
\be O\simeq \left( \begin{array}{ccc}
\frac{M_N}{\sqrt{M_N^2+m_D^2}}&
\frac{1}{\sqrt{2}}\frac{m_D}{\sqrt{M_N^2+m_D^2}}
+\frac{3}{4\sqrt{2}}\frac{M_N^2 m_D \mu_{S_2}}{(M_N^2+m_D^2)^2}&
\frac{1}{\sqrt{2}}\frac{m_D}{\sqrt{M_N^2+m_D^2}}
-\frac{3}{4\sqrt{2}}\frac{M_N^2 m_D \mu_{S_2}}{(M_N^2+m_D^2)^2} \\
\frac{M_N m_D \mu_{S_2}}{(M_N^2+m_D^2)^{3/2}}&
-\frac{1}{\sqrt{2}}-\frac{1}{4\sqrt{2}}\frac{M_N^2
\mu_{S_2}}{(M_N^2+m_D^2)^{3/2}}&
\frac{1}{\sqrt{2}}-\frac{1}{4\sqrt{2}}\frac{M_N^2 \mu_{S_2}}{(M_N^2+m_D^2)^{3/2}} \\
-\frac{m_D}{\sqrt{M_N^2+m_D^2}}&
\frac{1}{\sqrt{2}}\frac{M_N}{\sqrt{M_N^2+m_D^2}}
-\frac{1}{4\sqrt{2}}\frac{M_N(M_N^2+4
m_D^2)\mu_{S_2}}{(M_N^2+m_D^2)^2}&
\frac{1}{\sqrt{2}}\frac{M_N}{\sqrt{M_N^2+m_D^2}}
+\frac{1}{4\sqrt{2}}\frac{M_N(M_N^2+4 m_D^2)\mu_{S_2}}{(M_N^2+m_D^2)^2}\\
\end{array}\right).
\ee
Thus, the light neutrino mass can be of order eV, as required by
the oscillation data, for a {\rm TeV} scale $M_N$, provided $\mu_{S_2}$ is
sufficiently small, $\mu_{S_2} \ll M_N$. In this case, there is no any
restriction imposed on the value of Dirac mass $m_D$. Therefore,
the possibility of testing this type of model in LHC is quite
feasible. Note that in the limit $\mu_{S_2} \to 0$ which corresponds
to the unbroken $(-1)^{L+S}$ symmetry, we have massless light
neutrinos. Therefore, a small non-vanishing $\mu_{S_2}$ can be
considered as a slight breaking of a this global symmetry. Hence,
according to 't Hooft criteria, the smallness of $\mu_{S_2}$ is
natural. The possibility of generating small $\mu_{S_2}$ radiatively
has been discussed in Ref. \cite{Ma:2009gu}.

Finally, it is worth mentioning that the light neutrinos $\nu_l$
have suppressed mixing (of order $m_D \mu_{S_2}/(M_N^2 +m_D^2)$) with
one type of the heavy neutrinos (say $\nu_{H'}$) and
a rather small
mixing (of order $m_D/M_N$) with the other type of heavy neutrinos
($\nu_H$) by choosing appropriate parameters.
The mixing between the heavy neutrino $\nu_H$ and
$\nu_H'$ is maximal. In general, the physical neutrino states are
given
in terms of $\nu_L$, $N^c$, and $S_2$ as follows:%
\bea %
\nu_{l} &\simeq& \frac{M_N}{\sqrt{M_N^2+m_D^2}}\nu_L +\frac{M_N
m_D \mu_{S_2}}{(M_N^2+m_D^2)^{3/2}}N^c
-\frac{m_D}{\sqrt{M_N^2+m_D^2}}S_2
\simeq \nu_L+a_1 N^c-a_2 S_2\\
\nu_{H} &\simeq&
-\frac{1}{\sqrt{2}}\frac{m_D}{\sqrt{M_N^2+m_D^2}}\nu_L+\frac{1}{\sqrt2}~
N^c -  \frac{1}{\sqrt2}\frac{M_N}{\sqrt{M_N^2+m_D^2}}~S_2 \simeq
\alpha(- a_2 ~\nu_L + ~ N^c - ~S_2) \\
\nu_{H'} &\simeq&\frac{1}{\sqrt
2}\frac{m_D}{\sqrt{M_N^2+m_D^2}}\nu_L+ \frac{1}{\sqrt
2}N^c+\frac{1}{\sqrt2}\frac{M_N}{\sqrt{M_N^2+m_D^2}}~S_2
\simeq  \alpha(a_2 \nu_L+ ~N^c +  ~S_2).%
\eea %
For $m_D \simeq 100$ {\rm GeV}, $M_N \simeq 1$ {\rm TeV} and $\mu_{S_2} \simeq 1$
{\rm KeV}, one finds that $a_{1} \sim {\cal O}(10^{-10})$, $a_{2} \sim
{\cal O}(0.1)$, $a_3 \sim {\cal O}(0.07)$ and $\alpha \sim
\sin\pi/4$. In this respect, the gauge eigenstates for neutrinos
can be expressed in terms of the mass eigenstates as follow:

\begin{equation}
\left( \begin{array}{c}
\nu_L \\
 N^c \\
 S_2\\
\end{array}\right)
=O\left( \begin{array}{c}
\nu_l \\
 \nu_H \\
 \nu_{H'}\\
\end{array}\right)
\simeq \left( \begin{array}{ccc}
1&\alpha a_2&\alpha a_2\\
a_1&- \alpha& \alpha\\
- a_2& \alpha&\alpha\\
\end{array}\right)
\left( \begin{array}{c}
\nu_l \\
 \nu_H \\
 \nu_{H'}\\
\end{array}\right).
\end{equation}


\section{ $B-L$ Right-handed Sneutrino}

In our model, the sneutrino mass matrix of one generation is given
by $8\times 8$ matrix, which can be decomposed into the following
two mass matrices:  $(i)$ $6\times 6$ mass matrix in basis of
$(\tilde\nu_L, \tilde\nu^{\dagger}_L,\tilde N,\tilde N^{\dagger},
\tilde S_2, \tilde S^{\dagger}_2)^T$. $(ii)$ $2\times 2$ mass
matrix in basis of ($\tilde S_1, \tilde S^{\dagger}_1)^T$.
The $\tilde S_1$'s are decoupled and have now interactions with the SM
particles. Therefore, one can neglect it and focus on the $6\times
6$ sneutrino mass matrix. In the flavor basis; $\tilde\nu \equiv
(\tilde\nu_L, \tilde \nu^{\dagger}_L,\tilde N,\tilde N^{\dagger},
\tilde S_2, \tilde S^{\dagger}_2)^T$, the sneutrino mass matrix
$\tilde M^2_{\tilde\nu}$ is given as {\small
\begin{equation}
M^2_{\tilde{\nu}} = \left( \begin{array}{cccccc}
 M^2_{\tilde \nu^{\dagger}_L\tilde\nu_L}  &  0  &
 (M^2_{\tilde N^{\dagger}\tilde\nu_L})^{\dagger}  &  0  &
 (M^2_{\tilde S^{\dagger}_2\tilde\nu_L})^{\dagger}  &  0     \\
 0  &   (M^2_{\tilde \nu^{\dagger}_L\tilde\nu_L})^T  &
 0  &   (M^2_{\tilde N^{\dagger}\tilde\nu_L})^T  &  0  &
 (M^2_{\tilde S^{\dagger}_2\tilde\nu_L})^T         \\
 M^2_{\tilde N^{\dagger}\tilde\nu_L}  &  0  &
 M^2_{\tilde N^{\dagger}\tilde N}  & 0  &
 (M^2_{\tilde S_2^{\dagger}\tilde N})^{\dagger}  &
 (M^2_{\tilde S_2\tilde N})^{\dagger}  \\
 0  &  (M^2_{\tilde N^{\dagger}\tilde\nu_L})^{*}  &
 0  &
 (M^2_{\tilde N^{\dagger}\tilde N})^T  &
 (M^2_{\tilde S_2\tilde N})^{T}  &
 (M^2_{\tilde S_2^{\dagger}\tilde N})^{T}  \\
 M^2_{\tilde S^{\dagger}_2\tilde\nu_L}   &  0  &
 M^2_{\tilde S_2^{\dagger}\tilde N}  &
 (M^2_{\tilde S_2\tilde N})^{*}   &
 M^2_{\tilde S^{\dagger}_2\tilde S_2}  &
 (M^2_{\tilde S_2\tilde S_2})^{\dagger}   \\
 0  &  (M^2_{\tilde S^{\dagger}_2\tilde\nu_L})^{*}  &
 M^2_{\tilde S_2\tilde N}   &
 (M^2_{\tilde S_2^{\dagger}\tilde N})^{*}  &
 M^2_{\tilde S_2\tilde S_2}  &
 (M^2_{\tilde S^{\dagger}_2\tilde S_2})^T
\\
\end{array}\right),
\end{equation}
} where \bea M^2_{\tilde \nu^{\dagger}_L\tilde\nu_L}&=&\tilde
m^2_{\nu_L} +v^2\cos^2\beta Y^{e\dagger}Y^e+M^{\dagger}_DM_D +
\frac{m^2_Z}{2}\cos2\beta-\frac{M^2_{Z_{B-L}}}{4}(1-\cot^2\theta),\\
M^2_{\tilde N^{\dagger}\tilde N} &=&\tilde m^2_{N}
+v'^2\sin^2\theta Y^{S}Y^{S\dagger}+M_DM^{\dagger}_D -
\frac{M^2_{Z_{B-L}}}{4}(1-\cot^2\theta),\\
M^2_{\tilde S^{\dagger}_2 \tilde S_2}&=&\tilde
m^2_{S_2}+|\mu_{S_2}|^2+M^{\dagger}_{N}M_N -
\frac{M^2_{Z_{B-L}}}{2}(1-\cot^2\theta) ,\\
M^2_{\tilde N^{\dagger}\tilde \nu_L}&=&\mu^*v\cos\beta
Y^{\nu}+v\sin\theta Y^{\nu}_A,\quad
M^2_{\tilde S^{\dagger}_2\tilde\nu_L}=vv'\sin\theta\sin\beta
Y^{S\dagger}
Y^{\nu},\\
M^2_{\tilde S_2^{\dagger}\tilde N}&=&\mu' v'\cos\theta
Y^{S\dagger}+v'\sin\theta Y^{\nu\dagger}_A,\quad
M^2_{\tilde S_2 \tilde S_2}=B'_2\mu_{S_2},\quad M^2_{\tilde
S_2\tilde N}=\mu_{S_2}v'\sin\theta Y^{\dagger}_S.
\eea
In the case of existing the mixing of $\tilde{S}_2\tilde{S}_2$ or
$\tilde{S}_2\tilde{N}$, the complex scalar DM splits in two
real scalar and the lighter one is DM. If the mass split is small as well
as the momentum of DM, inelastic scattering can be considered.

The mass matrix is diagonalized by unitary matrix $\Gamma$ as
\begin{equation}
\Gamma^{\dag} M^2_{\tilde \nu} \Gamma={\rm diag}(m^2_{\tilde
\nu^m_1},m^2_{\tilde \nu^m_2},m^2_{\tilde \nu^m_3},m^2_{\tilde
\nu^m_4},m^2_{\tilde \nu^m_5},m^2_{\tilde \nu^m_6}).
\label{Utilde}
\end{equation}
Thus, the mass eigenstates $\tilde \nu^m$ are defined as $\tilde
\nu_i=\Gamma_{ij}\tilde \nu^m_j$. In general, the lightest
sneutrino can be written as a linear combination of the sneutrino
mass eigenstate, as follows:
\begin{equation}
 \tilde\nu^m_1 =  \Gamma^{\dag}_{11}\tilde\nu_L+  \Gamma^{\dag}_{12} \tilde\nu^{\dagger}_L+  \Gamma^{\dag}_{13}\tilde N +
  \Gamma^{\dag}_{14}\tilde N^{\dagger}+  \Gamma^{\dag}_{15}\tilde S_2+  \Gamma^{\dag}_{16}\tilde S^{\dagger}_2.
\end{equation}
\begin{figure}[ht]
\unitlength=1mm \hspace{-3cm}
\begin{picture}(70,80)
\includegraphics[height=8cm,width=12cm]{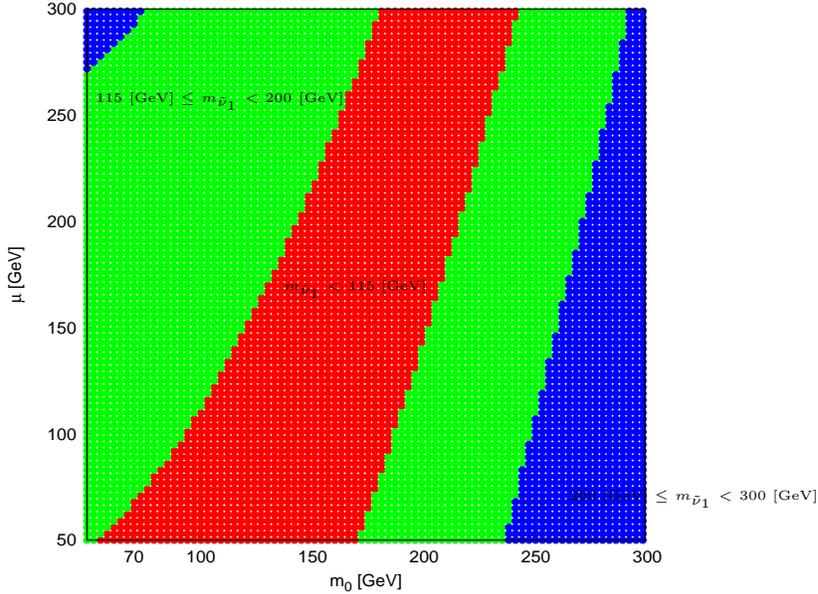}
\put(-108,65){ {\tiny 115 [{\rm GeV}] $\le$ $m_{\tilde\nu_1}$ $<$ 200 [{\rm GeV}] }}
\put(-83,40){ {\tiny $m_{\tilde\nu_1}$ $<$ 115 [{\rm GeV}] }}
\put(-45,12){ {\tiny 200 [{\rm GeV}] $\le$ $m_{\tilde\nu_1}$ $<$ 300 [{\rm GeV}] }}
\end{picture}
\caption{A contour plot for the mass of $B-L$ right-handed
sneutrino in terms of the most relevant parameters: soft mass term
$m_0$ and $\mu$ term and for $\theta\simeq \beta\simeq\pi/4$ and $Y_{\nu}\simeq 1$. The red region represents $m_{\tilde\nu_1}<$ 115 ${\rm GeV}$, which is kinematically excluded because DM mass is lighter than the SM Higgs mass. The green region represents 115 ${\rm GeV}\le$$m_{\tilde\nu_1}<$ 200 ${\rm GeV}$.  The blue region represents 200 ${\rm GeV}\le$$m_{\tilde\nu_1}<$ 300 ${\rm GeV}$. }
\label{dcontour}
\end{figure}

However if one considers large $\tan \beta$ and small $\tan
\theta$ limits, one finds that the lightest sneutrino is mainly
obtained from the
$(\tilde N, \tilde{S}_2)$ sector. Therefore, it can be expressed as %
\begin{equation}
 \tilde\nu_1 \simeq
  \Gamma^{\dag}_{13}\tilde N + \Gamma^{\dag}_{14}\tilde N^{\dagger}+
    \Gamma^{\dag}_{15}\tilde S_2+  \Gamma^{\dag}_{16}\tilde S^\dag_2.
\end{equation}
 The mass of this particle, which we call ``$B-L$ right-handed sneutrino" depends on the universal soft
 scalar mass $m_0$ and on the parameters $\mu$ and $\mu'$.
 In Fig. \ref{dcontour}, we display the mass range of $B-L$ right-handed sneutrino as
 function of the most relevant parameters $m_0$ and $\mu$.
 If this particle is the lightest SUSY particle, then it
is stable and can be considered as an interesting candidate for
DM.

Now we consider the relevant interactions of the $B-L$
right-handed sneutrino. From the superpotential ${\cal W}$ in Eq.
(\ref{sp}) one gets the following interacting Lagrangian of
$\tilde N$ in the flavor basis:
\bea
 {\cal L}^W_{int} &=& Y_{\nu ij}\tilde
N^{\dagger}_i\left[{(\tilde H^{0}_2)^c}P_L\nu_{Lj}-{(\tilde H^{+}_2)^c}P_L\ell^-_{Lj}\right]
+ Y_{Sij} \tilde N^{\dagger}_i
{(\tilde S_{2j})^c}P_L\widetilde\chi_1  
\nn\\&+& Y^{A}_{\nu ij}\tilde
N^{\dagger}_i\left(\tilde{\nu}_{Lj}H^0_2-\tilde{\ell}^-_{Lj}H^+_2\right)
+Y^{A}_{S ij}\tilde N^{\dagger}_i\tilde S_{2j}\chi_1+h.c.,
\eea


Also from the $F$-term contributions to the scalar potential one
finds the following interaction terms:%
\bea {\cal L}^F_{int} &=&
-|H^-_1|^2\tilde\nu^{\dagger}_{Li}(Y^{\dagger}_eY_e)_{ij}\tilde
N_j - (|H^+_2|^2+|H^0_2|^2)\tilde N^{\dagger}_{i}(Y^{\dagger}_\nu
Y_\nu)_{ij}\tilde N_j \nn\\&-&
(H^{0\dagger}_1H^+_2+H^{-\dagger}_1H^0_2)\tilde
N^{\dagger}_{i}(Y^{\nu}Y^{e\dagger})_{ij}\tilde \ell_{Rj} -
(\tilde N^{\dagger} Y_{\nu}\tilde \nu_L)(\tilde\nu^{\dagger}_L
Y^{\dagger}_\nu\tilde N) -\mu^*H^{0\dagger}_1(\tilde N^{\dagger}
Y_{\nu}\tilde \nu_L)
\nn\\
&-& (\tilde N^{\dagger} Y_{\nu}\tilde
\ell_L)(\tilde\ell^{\dagger}_L Y^{\dagger}_\nu\tilde N)
-\mu^*H^{-\dagger}_1(\tilde N^{\dagger} Y_{\nu}\tilde \ell_L) -
(\tilde S^{\dagger}_2 Y^{\dagger}_S\tilde N)(\tilde N^{\dagger}
Y_{S}\tilde S_2) - \mu^*\chi^{\dagger}_2(\tilde
N^{\dagger}Y_{S}\tilde S_2) \nn\\&-& |\chi_1|^2\tilde
N^{\dagger}_i(Y_SY^{\dagger}_{S})\tilde N_j
-\mu^*_{S_2}\chi_1\tilde N^{\dagger}_iY_{S ij}\tilde
S^{\dagger}_{2j}+h.c.. \eea

Next the interactions of $\tilde{N}$ with the gauge fields lead to
the following Lagrangian:
\be {\cal L}^G_{int} = g^2_{B-L}Z^2_{B-L}\tilde
N^{\dagger}_i\tilde N_i -ig_{B-L}Z^{\mu}_{B-L}\left(\tilde
N_i\partial_{\mu}\tilde N^{\dagger}_i-\tilde
N^{\dagger}_i\partial_{\mu}\tilde N_i  \right) -
i\sqrt2g_{B-L}\tilde N_{i} N^c_i\tilde Z_{B-L} +h.c.. \ee

Finally, the $D$-term implies that
\bea {\cal L}^D_{int} &=& g^2_{B-L}(\tilde N^{\dagger}\tilde N)\nn\\
&\times&
\left( -|\tilde\nu_L|^2 -|\tilde N|^2
 +\frac{|\tilde u_L|^2}3+\frac{|\tilde d_L|^2}3+\frac{|\tilde u_R|^2}3+\frac{|\tilde d_R|^2}3
 -|\tilde \ell_L|^2-|\tilde \ell_R|^2
 +2|\tilde S_1|^2-2|\tilde S_2|^2+|\tilde \chi_1|^2-|\tilde\chi_2|^2\right) .\nn\\
\eea

 In our analysis for $B-L$ right-handed sneutrino
($\tilde \nu_1$) DM, the following assumptions are considered for
simplification: $(i)$ Each sector consists of one generation only.
$(ii)$ $m_{SM}$ $<$ $m_{DM}$ $<$ $m_{SUSY}$, $m_{h^\pm}$,
$m_{\chi_{1,2}}$, $M_{Z_{B-L}}$, $m_{S_{1,2}}$, where $m_{SM}$ is
standard model particles including of the lightest neutral Higgs
boson ($h_0$), $m_{SUSY}$ is the supersymmetric particles, and
$m_{DM}$ is the mass of the lightest SUSY sneutrinos. $(iii)$ In
chargino sector, $h^\pm$ mass is approximately given by $\mu$, in
the limit of $\mu$, $M_2$ $>>$ $m_W$, where $m_W$ is the SM
charged weak gauge boson mass. Under these assumptions, the
relevant interacting Lagrangian of $\tilde\nu_1$ consists of \be
{\cal L}_{int}={\cal L}^W_{int}+{\cal L}^F_{int}+{\cal L}^G_{int}.
\ee%
Note that the D-term is kinematically irrelevant, while the $W, F,
G$ interactions take the following simplified form:
\bea %
{\cal L}^W_{int} &=&
 Y'_{\nu}\Gamma_{41}\tilde\nu^{\dag}_1\left[\overline{\tilde h^{0c}}P_L\nu_{L}-\overline{\tilde h^{+c}}P_L\ell^-_{L}\right]+h.c.,\label{pamela}
\\
{\cal L}^F_{int} &=& -(Y^{\dagger}_\nu Y_\nu)
\Gamma_{41}\Gamma_{31}|h^0|^2\tilde\nu^{\dagger}_1\tilde\nu_1
,\label{wmap}
\\
{\cal L}^G_{int} &=& -ig_{B-L}Z^{\mu}_{B-L}\Gamma_{41}\Gamma_{31}\left(\tilde\nu_1\partial_{\mu}\tilde\nu^{\dagger}_1-\tilde\nu^{\dagger}_1\partial_{\mu}\tilde\nu_1  \right) ,\label{direct-dt}
\ee 
where $Y'_{\nu}$ is considered to be included in the mixing of
Higgsino and Chargino. Hereafter we use it as $Y_{\nu}$. Eq.
(\ref{pamela}) might be relevant to the indirect detection, which
will be discussed in the section \ref{indt}. Eq. (\ref{wmap}) is
more relevant to the WMAP experiment, which will be discussed in
the next section. Eq. (\ref{direct-dt}) is applied to analyze the
direct detection as CDMS II/XENON, which will be also discussed in the
section \ref{dt}.

\section{Relic abundance of $B-L$ right-handed sneutrino }
In this section, we compute the relic abundance of $B-L$
right-handed sneutrino DM. We consider the standard computation of
the cosmological abundance, where $\tilde \nu_1$ is assumed to be
in thermal equilibrium with the SM particles in the early universe
and decoupled when it was non-relativistic. Therefore, the $\tilde
\nu_1$ density can be obtained by solving the Boltzmann equation:
\be \frac{d
n_{\tilde{\nu}_1}}{dt}+3Hn_{\tilde{\nu}_1}=-<\sigma^{ann}_{\tilde{\nu}_1}v>
[(n_{\tilde{\nu}_1})^2-(n^{eq.}_{\tilde{\nu}_1})^2],\label{boltzmann}
\ee %
where $n_{\tilde{\nu}_1}$ is $\tilde{\nu}_1$ number density with
$\rho_{\tilde{\nu}_1}=m_{\tilde{\nu}_1}n_{\tilde{\nu}_1}$. One
usually defines
$\Omega_{\tilde{\nu}_1}=\rho_{\tilde{\nu}_1}/\rho_{c}$, where
$\rho_{c}$ is the critical mass density. In addition,
$<\sigma^{ann}_{\tilde{\nu}_1}v>$ is the thermal averaged of the
total cross section for $\tilde{\nu}_1$ annihilation into SM
lighter particles times the DM relative velocity $v$. For non-relativistic
$\tilde{\nu}_1$, the thermal averaged annihilation cross section,
$<\sigma^{ann}_{\tilde{\nu}_1}v>$, can
be approximated as follows \cite{griest1}: %
\be %
<\sigma^{ann}_{\tilde{\nu}_1} v> \simeq  a_{\tilde{\nu}_1} + b_{\tilde{\nu}_1} v^2, %
\ee %
where $a_{\tilde{\nu}_1}$ $b_{\tilde{\nu}_1}$ are the
coefficients coming from $s$-wave and $p$-wave of
$\tilde{\nu}_1\tilde{\nu}_1$ annihilation, respectively.

\begin{figure}[htb]
\unitlength=1mm \hspace{-7cm}
\begin{picture}(35,40)
\includegraphics[height=3.6cm]{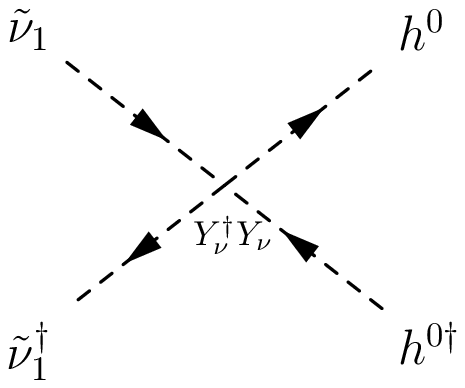}
\hspace{2cm}
\includegraphics[height=3.6cm]{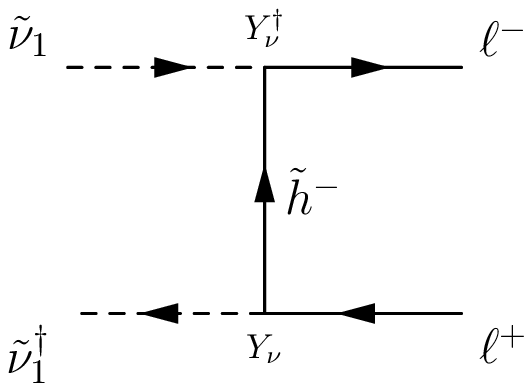}
\end{picture}
\caption{Possible annihilation channels of $\tilde{\nu}_1$. The
second diagram gives a sub-dominant contribution, however it may
be relevant for indirect detection processes.}
\label{wmap-diagram}
\end{figure}

From Eqs.(\ref{pamela}-\ref{direct-dt}), one finds that the
dominant annihilation channels of $\tilde{\nu}_1$ are given in
Fig. \ref{wmap-diagram}. It turns out the annihilation of $B-L$
right-handed sneutrino into SM-like neutral Higgs, through the
four point interaction vertex, gives the dominant contribution.
The tree level annihilation channel $\tilde{\nu}_1 \tilde{\nu}_1
\to \ell^+ \ell^-$ is suppressed by the mass of the chargino exchanged
particle. This channel may be relevant for the indirect detection
processes which will be discussed later. Our computation for the
annihilation cross section leads to the following
$a_{\tilde{\nu}_1}$ $b_{\tilde{\nu}_1}$: %
\bea %
a_{\tilde{\nu}_1}&=& \frac{\beta'_{h^0}}{32\pi m^2_{\tilde
\nu_1}}|Y_{\nu}  \Gamma_{31} \Gamma_{41}  |^4,\\
b_{\tilde{\nu}_1}&=&  \frac{\beta'_{h^0}(x^2_{h^0}-1)}{128\pi
m^2_{\tilde \nu_1}}|Y_{\nu} \Gamma_{31} \Gamma_{41}|^4,\label{af-bf} %
\eea%
where %
\be %
z_a=\frac{m_a}{m_{\tilde{\nu}_1}},\quad
\beta'^2_{a}=1-z^2_a,~x^2_a=\frac{z^2_a}{2(1-z^2_a)},~\label{dfnt}
\ee %
here we define that $a$ is a final-state particle.
From Eq.(\ref{boltzmann}) one finds that the relic abundance
$\Omega_{\tilde{\nu}_1} h^2$ is given by \cite{Griest:1989zh} %
\be %
\Omega_{\tilde{\nu}_1} h^2 \simeq \frac{8.76\times
10^{-11}{\rm GeV}^{-2}}{g^{1/2}_{*}(T_F)(a_{\tilde{\nu}_1}/x_F+3b_{\tilde{\nu}_1}/x^2_F)},
\label{relic-abundance}%
\ee%
where %
\be%
x_F=\ln\frac{0.0955~ m_{\mathrm{pl}}~
m_{\tilde{\nu}_1}(a_{\tilde{\nu}_1}+6b_{\tilde{\nu}_1}/x_F)}{(g^{1/2}_{*}(T_F)
x_F)^{1/2}}. %
\ee %
Here $m_{{\rm pl}}$ is the Planck mass ($1.22\times10^{19}$ {\rm GeV})
and $g_{*} (T_F)$ enumerates the degrees of freedom of
relativistic particles at the freeze out temperature $T_F$, which
can be fixed as $g^{1/2}_{*} (T_F)=10$. From the above
expressions, one notes that $\tilde{\nu}_1$ relic abundance
depends only on the right-sneutrino mass and the annihilation
cross section coefficients $a_{\tilde{\nu}_1}$ and
$b_{\tilde{\nu}_1}$.

\begin{figure}[htb]
\unitlength=1mm \hspace{-19cm}
\begin{picture}(10,45)
\includegraphics[height=5cm,width=6cm]{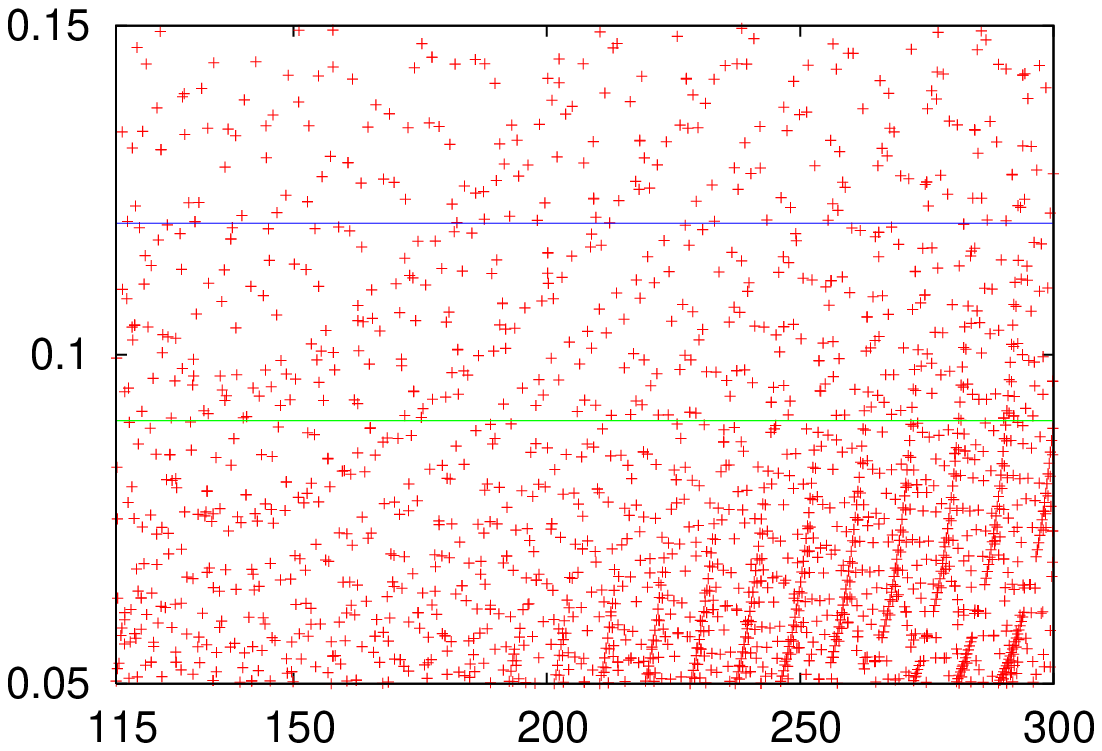}~~
\includegraphics[height=5cm,width=6cm]{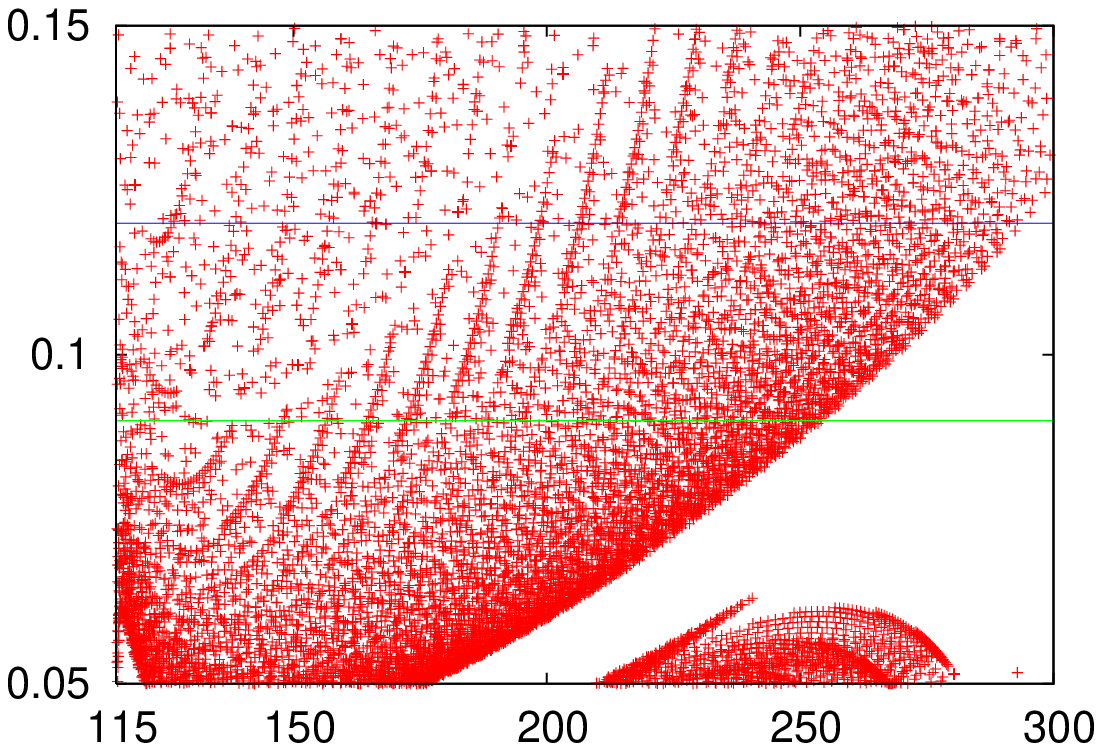}~~
\includegraphics[height=5cm,width=6cm]{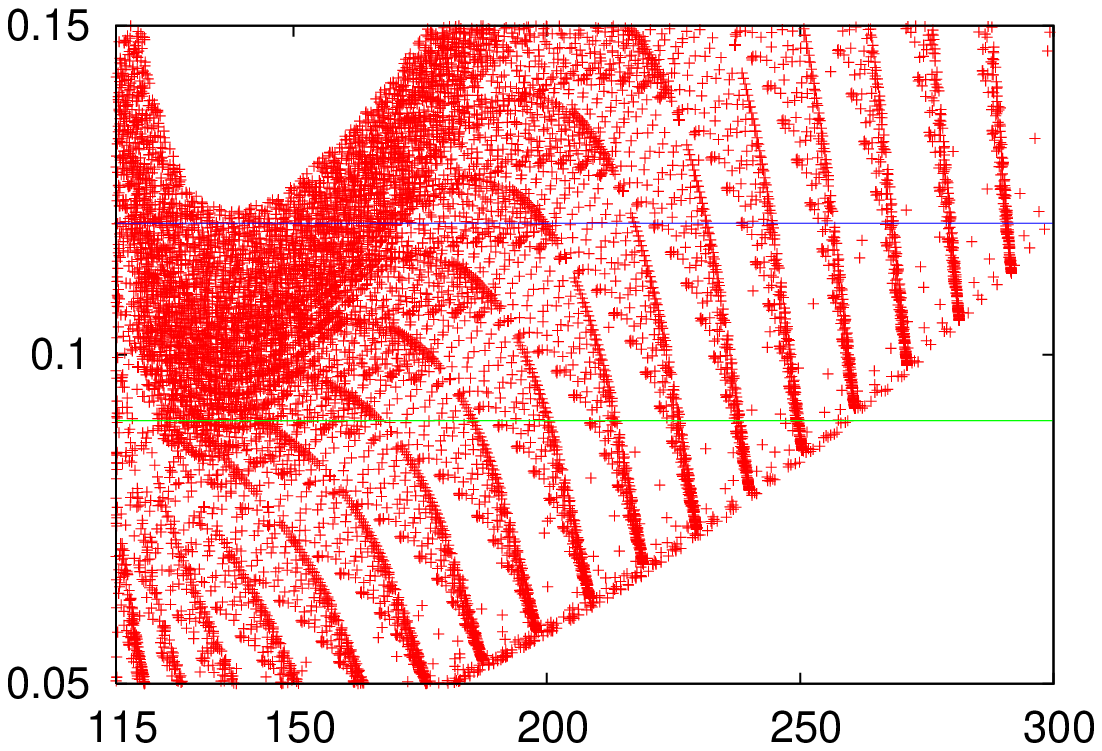}
\put(-45,2.5){ $m_{\tilde\nu_1}$ [{\rm GeV}] ($Y_{\nu}=0.5$)}
\put(-105,2.5){ $m_{\tilde\nu_1}$ [{\rm GeV}] ($Y_{\nu}=0.7$)}
\put(-170,2.5){ $m_{\tilde\nu_1}$ [{\rm GeV}] ($Y_{\nu}=1.0$)}
\put(-125,30){ $\Omega_{\tilde\nu_1}h^2$ }
\end{picture}
\caption{Relic density $\Omega_{\tilde{\nu}_1} h^2$ as function of
the $B-L$ right-handed sneutrino mass $m_{\tilde{\nu}_1}$ for
three values of Dirac neutrino Yukawa coupling. The region between
two lines is allowed by the experiment of WMAP
\cite{Spergel:2006hy}. } \label{relic}
\end{figure}

Given $a_{\tilde{\nu}_1}$ $b_{\tilde{\nu}_1}$ we can determine the
freeze out temperature $T_F$, below which the $B-L$ right-handed
sneutrino annihilation rate is smaller than the expansion rate of
the universe and then computing the relic density
$\Omega_{\tilde{\nu}_1} h^2$. In our numerical computation we
assume a universal soft mass ($m_0$) and fix the
other parameter as follows: %
\bea %
&& m_{h^0}=115~ {\rm GeV},~M_{Z_{B-L}}= 6~ {\rm TeV}, \quad M_N=
1~ {\rm TeV}, ~ \mu_{S_{1,2}} = 1~ {\rm KeV},  ~ v = 175~{\rm
GeV}, \nn\\&& m_D =100~ {\rm GeV},~B'_{1,2}=v' =\mu^{(')}=1~ {\rm
TeV},\quad Y_S=0.1, ~Y^{e}=0.05,~Y^A_{\nu}=Y^A_{S}=0.1~ {\rm
TeV}, \eea
In addition we  analyze the relic density of $\tilde{\nu}_1$ in
the following regions of the parameter $\mu$, $\beta$, and
$\theta$:
 \be
&&0< \theta~,~\beta <\pi,~~~~~~
 50\ {\rm GeV}< \mu~,~m_0< 300\ {\rm GeV}.
 \ee

In Fig. \ref{relic}, we present the values of relic abundance of
$B-L$ right-handed sneutrino  $\Omega_{\tilde\nu_1} h^2$ as a
function of $m_{\tilde{\nu}_1}$ for the following values of Dirac
neutrino Yukawa coupling: $Y_{\nu}=0.5,0.7, 1.0$. We require $B-L$
right-handed sneutrino relic density to be $0.09 <
\Omega_{\tilde\nu_1} h^2< 0.12$ in order to be consistent with
WMAP results at $3\sigma$ \cite{Antoniadis:2010nb}. As can be seen
from this figure, smaller values of Dirac neutrino Yukawa coupling
are favored and lead to more allowed points that satisfy the WMAP
observational limits of DM relic density. Also we find that the
smaller $Y_{\nu}$ is considered, the smaller DM mass one obtains.

\section{ Direct detection \label{dt}}

In this section we discuss the possibility to detect our $B-L$
right-handed sneutrino in direct detection experiments such as
CDMS (II) \cite{Ahmed:2009zw} and XENON 100 experiment
\cite{Aprile:2010um}. The general form of the elastic scattering
cross section between DM $\tilde\nu_1$ and nuclei $N$
is given by \cite{Goodman:1984dc, griest1} %
\begin{figure}[b]
\unitlength=1mm \hspace{-1cm}
\begin{picture}(35,40)
\includegraphics[width=6cm]{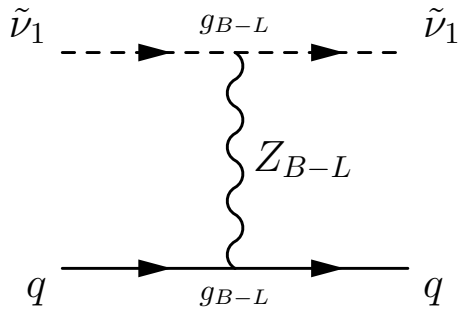}
\end{picture}
\caption{Our dominant diagram for the direct detection }
\label{drt-dia}
\end{figure}
\be%
\sigma^{\mathrm{vec}}_{\tilde\nu_1-N} = \frac{m_r^2}{16\pi}\left|b_N\right|^2. %
\ee%
The reduced mass $m_r$ is defined as
\be%
m_r=\left(\frac{1}{m_{\tilde\nu_1}}+\frac{1}{M}\right)^{-1}, %
\ee %
where $M$ is the nuclei mass. The coefficient $b_N$ is given by%
\bea %
b_N=(A-Z)b_n+Zb_p, \quad b_p=2b_u+b_d,\quad b_n=b_u+2b_d, %
\eea %
Here $A$ and $Z$ are the mass number and the atomic number, respectively.
The effective Lagrangian parameters $b_u$ and $b_d$ are defined as
\begin{equation}
\mathcal{L}_{\mathrm{eff}}= b_q X^\mu \bar{q}\gamma^\mu q, \quad
q=(u~,~d).
\end{equation}
Here $X^\mu$ is a general form of the vector current. In case of
fermionic DM $X^\mu$ is given by $X^\mu\simeq\bar f\gamma^\mu f$.
While for bosonic DM, it is defined as $X^\mu\simeq ib^\dag
\partial^\mu b-ib\partial^\mu b^\dag$.


In our $B-L$ case, the elastic scattering cross section of the
right-handed sneutrino with a given nuclei has a spin-independent
contribution arising from $Z_{B-L}$ gauge boson exchange diagrams,
as can been seen in Fig. \ref{drt-dia}.
The interactions between the right-handed sneutrino $\tilde\nu_1$
and $Z_{B-L}$ boson in $B-L$ model are \bea \mathcal{L}&\supset&
-ig_{B-L}Z^{\mu}_{B-L}\Gamma_{41}\Gamma_{31}\left(\tilde\nu_1\partial_{\mu}\tilde\nu^{\dagger}_1-\tilde\nu^{\dagger}_1\partial_{\mu}\tilde\nu_1  \right)
-\frac{1}{3}g_{B-L}Z_{B-L}^\mu\bar{u}\gamma_\mu u
-\frac{1}{3}g_{B-L}Z_{B-L}^\mu\bar{d}\gamma_\mu d. \eea
Writing down the effective interaction, one obtains%
\bea%
i\mathcal{L}_{\mathrm{eff}}\supset
-\frac{|\Gamma_{41}\Gamma_{31}|^2}{M_{Z_{B-L}}^2}\frac{g_{B-L}^2}{3}
\left(\tilde\nu_1\partial_{\mu}\tilde\nu^{\dagger}_1-\tilde\nu^{\dagger}_1\partial_{\mu}\tilde\nu_1
\right) \bar{u}\gamma^\mu u
-\frac{|\Gamma_{41}\Gamma_{31}|^2}{M_{Z_{B-L}}^2}\frac{g_{B-L}^2}{3}
\left(\tilde\nu_1\partial_{\mu}\tilde\nu^{\dagger}_1-\tilde\nu^{\dagger}_1\partial_{\mu}\tilde\nu_1
\right) \bar{d}\gamma^\mu d.%
\eea %
Assuming $|\Gamma_{41}\Gamma_{31}|^2\simeq1$ for simplicity, one
finds %
\be
b_p=b_n=i\frac{g_{B-L}^2}{M_{Z_{B-L}}^2}.%
\ee%
Therefore, $b_N$ is given by%
\be%
b_N=iA\frac{g_{B-L}^2}{M_{Z_{B-L}}^2}. %
\ee %
%
\begin{figure}[t]
\unitlength=1mm \hspace{-5cm}
\begin{picture}(35,55)
\includegraphics[height=6cm,width=8cm]{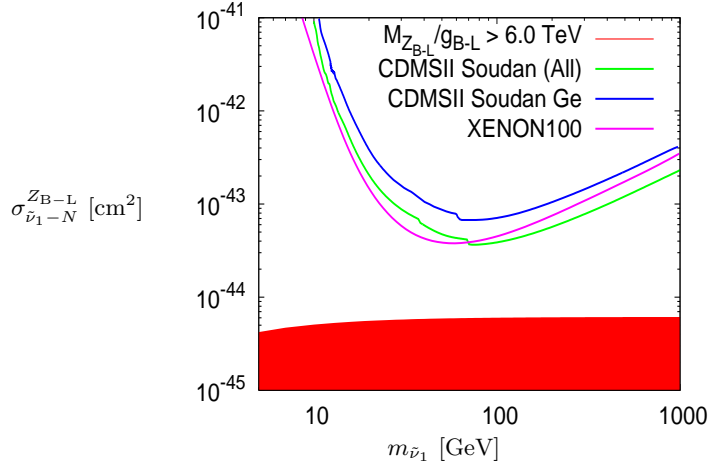}
\put(-45,-2.0){ $m_{\tilde\nu_1}$ [{\rm GeV}] }
\put(-95,30){ $\sigma^{Z_{{\rm B-L}}}_{\tilde\nu_1-N}\ [{\rm cm}^2]$ }
\end{picture}
\caption{Elastic scattering cross section between a nuclei and the
$B-L$ right-handed sneutrino DM as function of the DM mass for
$g_{B-L}=1$.} \label{drt-crs}
\end{figure}
%
Thus, the elastic scattering cross section of $B-L$ right-handed
sneutrino is given by%
\be
\sigma^{Z_{B-L}}_{\tilde\nu_1-N}=\frac{m_r^2}{16\pi}|A|^2\frac{g_{B-L}^4}{M_{Z_{B-L}}^4}.
\ee

In Fig. \ref{drt-crs} we depict the relation between the cross
section and $B-L$ right-sneutrino mass. As can been seen from this
figure, the following upper bound on
$\sigma^{Z_{B-L}}_{\tilde\nu_1-N}$ can be obtained:%
\be
\sigma^{Z_{B-L}}_{\tilde\nu_1-N}\le 6.2\times10^{-45}\ {\rm
cm^{2}}. %
\ee %
It is also remarkable that the elastic cross section is quite
insensitive to the $B-L$ right-handed sneutrino mass
$m_{\tilde\nu_1}$. However, one observes that a light sneutrino
$\sim {\cal O}(100)$ {\rm GeV} is more favored by direct detection
experimental results. The current limits from CDMS II and XENON
experiments indicate to a lower-bound of order $3.7\times10^{-44}$
${\rm cm}^{2}$. This suggests that our $B-L$ right-handed
sneutrino DM is expected to be detected in the direct detection
experiments in near future.

Before concluding this section, it is worth mentioning that 
the XENON 100 \cite{Aprile:2011ts} experiment has recently presented 
new limits on the WIMP-nucleon cross section for inelastic DM. 
These limits are due to a DM run with 100.9 live-days of data, 
taken from January to June 2010. It was shown that $\sigma_{\chi-N} < 10^{-41}$ ${\rm cm}^{2}$ 
can be extracted for $m_\chi > 100$ ${\rm GeV}$. Where $\chi$ is a generic dark matter. 
The bound rules out the explanation of controversial DAMA/LIBRA modulation results,  
as being due to inelastic DM.
The bound obtained on $\sigma_{\chi-N}$ from the inelastic DM analysis of XENON 100 
should be considered carefully, since a minimum velocity for DM to scatter in a detector is introduced, 
hence a large amount of fiducial is needed. 
Nevertheless, as can be seen from Eq. (52), the cross section of our $B-L$ sneutrino is well below the XENON 100 bound.

\section{ Indirect detection  \label{indt}}
\subsection{PAMELA and Fermi-LAT experiments}

As advocated in the introduction, the indirect searches for DM by
the Space Observatory PAMELA \cite{Adriani:2008zr} and Fermi-LAT
\cite{Abdo:2009zk}, indicate that the DM may contribute to the
positron flux by direct annihilation in $\ell^+ \ell^-$. This is one of
the main feature of our DM candidate $B-L$ right-handed sneutrino,
therefore, it important to investigate the possibility that it
accounts for these results.  PAMELA collaboration reported excess
flux between $8$ and $80$ {\rm GeV}, with no excess in the corresponding
anti-proton flux. Also ATIC and Fermi-LAT balloon experiments have
shown excess electron and positron flux at energies around $10-
1000$ {\rm GeV}. While there are plausible astrophysical explanations for
these excesses, such as local pulsars and supernovas remnants,
they could also result from DM annihilation. Note that in order to
explain the Fermi-LAT experiment by DM annihilation, the DM mass
must be of order ${\cal O}({\rm TeV})$. As shown in the previous
section, such heavy DM mass is not favored by of direct
detections. Therefore, in our analysis, we assume that the
Fermi-LAT experiment may be saturated by considering an
astrophysical background and we will focus on PAMELA measurement.

It is known that if the DM annihilation is to explain the observed
anomalous flux, a large annihilation cross section, $\langle
\sigma v\rangle\simeq 10^{-24}\ {\rm cm^3s^{-1}}$, is required to
fit the excess flux, which is incompatible with straightforward
estimates of the relic DM abundance in conventional cosmological
models. Otherwise, a huge, unexplained, boost factor must be
introduced. In our SUSY $B-L$ model, the $B-L$ right-handed
sneutrino annihilates into $\ell^+ \ell^-$ channels, as shown in the
second diagram of Fig. \ref{wmap-diagram}. However, as discussed
in the previous section that these channels give sub-dominate
contribution to the annihilation process. Therefore, the
corresponding annihilation cross section is $< 10^{-27} \ {\rm
cm^3s^{-1}}$. In this case, one requires a huge boost factor
$~{\cal O}(10^{5-6})$ at least in order to account for PAMELA
results. In general, it is known that there are two mechanisms to
enhance the cross section and may justify this large boost factor.
The first is Breit-Wigner mechanism \cite{bw} and the second is
the Sommer-feld \cite{sf} enhancement. Breit-Wigner mechanism can
not be implemented in our model, since there is no any diagrams
with $s$-channel. On the other hand,  the  Sommer-feld enhancement
requires higher DM mass compared to the mediated particles in
order to obtain enough large boost factor $>{\cal O}(10^{5-6})$.
This assumption we have already avoided in order to not spoil the
direct detection. As a result, it is difficult for our $B-L$
right-handed neutrino to explain the controversial results of
PAMELA experiment.

\subsection{Muon flux measurement from Super-Kamiokande}

The high energy neutrinos induced by DM annihilations in the
earth, the sun, and the galactic center is an important signal for
the indirect detection of DM. Such energetic neutrinos induce
upward through-going muons from charged current interactions
provide the most effective signatures in Super-Kamiokande. The
neutrino-induced muon flux is evaluated from the neutrino flux
\cite{Ritz:1987mh,Jungman:1995df} as 
\begin{equation}
    F_{\mu^+ \mu^-}^{(\rm ann)} \simeq 5.9\times 10^{-15}~{\rm cm^{-2}s^{-1}} 
\times \sum_F S_F
    \left ( \frac{\langle \sigma v\rangle _F }{10^{-23}~{\rm cm^3s^{-1}}} \right )
    \left ( \frac{\langle J_2 \rangle_\Omega \Delta \Omega}{10} \right ).  \
\label{muann}
\end{equation}
where we fix the typical values $\langle J_2 \rangle_\Omega \Delta
\Omega \sim 10$ for $\psi_{\rm max} =5^\circ$ in case of the
Navarro-Frenk-White (NFW) halo density profile
\cite{Navarro:1995iw}, and $F$ collectively denotes the primary annihilation mode (e.g.,
$\tau^+ \tau^-$, etc.). Notice that model dependence comes from
$\langle \sigma v\rangle _F$, which will be shown later.
 $S_F$ is defined as
\begin{equation}
    S_F = \sum_{\nu_i}\int_{E_{\rm min}}^{E_{\rm in}}\frac{dN_F^{(\nu_i)}}{dE}P_{\nu_i \nu_\mu}
    \left ( \frac{E}{E_{\rm in}} \right )^2 dE,
\end{equation}
where $E_{\rm in}=m_{\tilde\nu_1}$ , and $E_{\rm min}$ is the
threshold energy above which the muons can be detected.
$P_{\nu_i\nu_\mu}$ denotes the probability that the $\nu_i$ at the
production is observed as $\nu_\mu$ at the Earth due to the effect
of neutrino oscillation. Regardless to the complicated expression
of $S_F$, it is found as a fixed value depending on each of the
$F$ particle, which is, e.g., $0.2$ for $\mu$ pair, $0.14-0.18$
for $\tau$ pair, $0.78$ for $\nu_{\tau}$ pair, etc
\cite{Hisano:2008ah}. The limits from Super-Kamiokande are given
in the ref. \cite{Desai:2004pq}.


\begin{figure}[htb]
\unitlength=1mm \hspace{-1.0cm}
\begin{picture}(35,40)
\includegraphics[width=5.0cm]{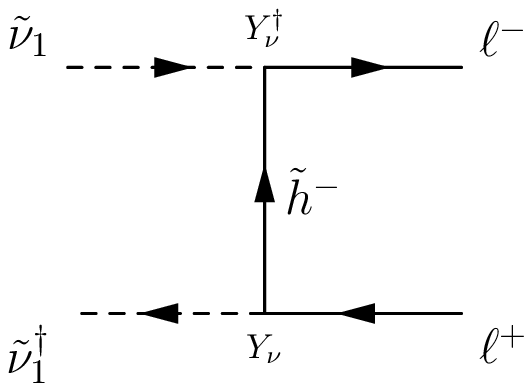}
\end{picture}
\caption{Fenymen diagram of $\tilde{\nu}_1 \tilde{\nu}_1 \to \ell^-
\ell^+$. Note that, in general, there is an another contribution of
$\tilde \nu_1,\tilde \nu_1\rightarrow \tilde h^0\rightarrow
\nu,\nu$ process. However, due to the masslless limit of $\nu$,
the cross section of this channel vanishes.}
\label{pamela-diagram}
\end{figure}
In SUSY $B-L$ model, the relevant Lagrangian in given by %
\bea %
{\cal L}^W_{int} = Y_{\nu}\Gamma_{41}\tilde\nu^{\dag}_1\left[\overline{\tilde h^{0c}}P_L\nu_{L}-
\overline{\tilde h^{+c}}P_L\ell^-_{L}\right]+h.c.
\eea
One can show that in the limit  $v\rightarrow0$, the thermally
averaged cross section is given by%
\be %
\langle\sigma
v\rangle_F|_{v\rightarrow 0}&\simeq& \sum_{F=\tau,\nu_{\tau}}
\frac{3\beta'_F}{16\pi
m^2_{\tilde\nu_1}}|Y_{\nu}\Gamma_{31}\Gamma_{41}
|^4z^2_F(2-z^2_F) \times\left(r^2_{F,\tilde h^+}+r^2_{F,\tilde
h^-}+r_{F,\tilde h^-}r_{F,\tilde h^+} +r^2_{F,\tilde h^0}
\right)\nn\\
&\simeq& \sum_{F=\tau} \frac{9\beta'_F}{16\pi
m^2_{\tilde\nu_1}}|Y_{\nu}\Gamma_{31}\Gamma_{41}|^4z^2_F(2-z^2_F) \times r^2_{F,\tilde h^+}. \label{aw} %
\ee%
Note that the term proportional to $r^2_{F,\tilde h^0}$ vanishes
in the massless limit of $\nu$. The parameters $z_F$ and
$\beta'_F$ are as defined in Eq. (\ref{dfnt}). The other
parameters are defined as follows:
\bea%
w_{\alpha}=\frac{m_{\alpha}}{m_{\tilde\nu_1}}, ~~~~~
r_{F\alpha}=(1-z^2_F+w^2_{\alpha})^{-1}. %
\eea %
Here $F$ refers to the final-state particle and $\alpha$ denotes
the mediated-particle. In our analysis for the muon flux, we
consider the same set of inputs that we have used in the previous
sections that leads to relic abundance within the WMAP limits:
$0.09 < \Omega h^2< 0.12$. Moreover, since we are considering the
effect one generation, we assume $\tau$ final state only.  
Using $S_F$ that corresponds to $\tau$ final state, one finds
$\langle\sigma v\rangle$ %
\be
F_{\mu^+ \mu^-}^{(\rm ann)} &\simeq& 5.9\times 10^{-15}~{\rm
cm^{-2}s^{-1}} \times \sum_{F=\tau^{\pm}} S_F \left( \frac{\langle
\sigma v\rangle _F }{8.56\times 10^{-7}} \right) \left(
\frac{\langle J_2 \rangle_\Omega \Delta \Omega}{10} \right)
\nn\\&\simeq& 6.9\times 10^{-9}~{\rm cm^{-2}s^{-1}} \times
\sum_{F=\tau^{\pm}} S_F \left (\frac{9\beta'_F}{16\pi m^2_{\tilde
\nu_1}}|Y_{\nu} \Gamma_{31}\Gamma_{41} |^4z^2_F(2-z^2_F) \times r^2_{F,\tilde
h^+} \right ).
\ee
As we fixed the cone-half angle from the galactic center:
$\psi_{\rm max} =5^\circ$ which is maximum in the case of NFW
profile, one finds that the Super-Kamiokande limit should be less
than $5\times 10^{-15}{\rm cm^{-2}sec^{-1}}$ (See figures of Ref.
\cite{Hisano:2008ah}).

\begin{figure}[htb]
\unitlength=1mm \hspace{-5.0cm}
\begin{picture}(55,60)
\includegraphics[height=6cm,width=10cm]{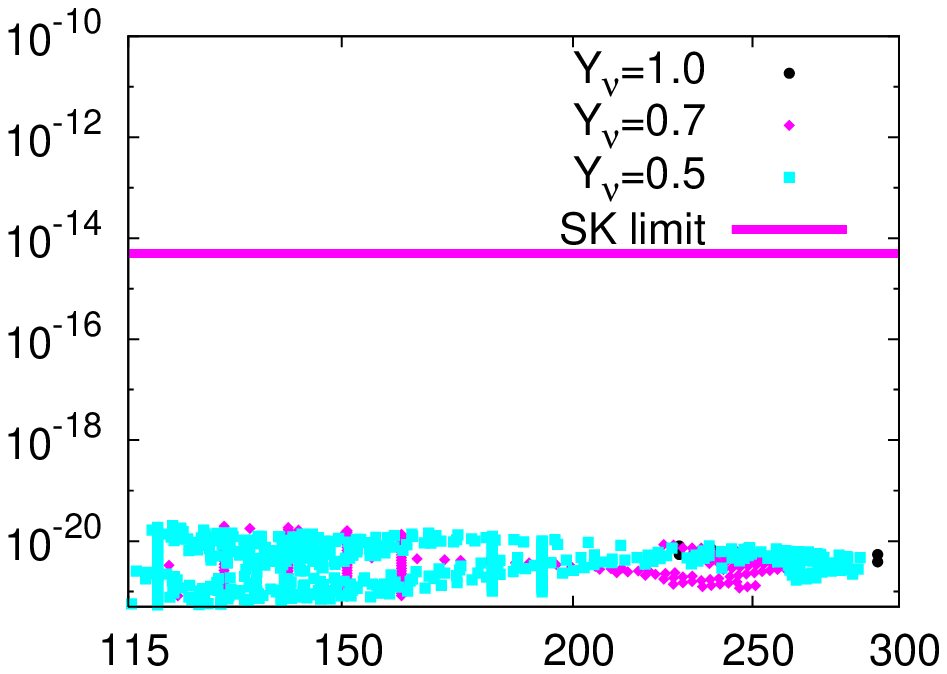}
\put(-43,0){ $m_{\tilde\nu_1}$ [{\rm GeV}] }
\put(-110,32){ $F_{\mu^+ \mu^-}^{(\rm ann)}\ [{\rm cm^{-2}sec^{-1}}]$ }
\end{picture}
\caption{Muon flux induced from the $B-L$ annihilation in the
galactic center. The horizontal line refers to the
Super-Kamiokande upper bound which is given by $\sim 5\times
10^{-15}{\rm cm^{-2}sec^{-1}}$. } \label{muon-flux}
\end{figure}

In Fig. \ref{muon-flux}, we plot the muon flux induced from the
$B-L$ annihilation in the galactic center. As can be seen from
this figure, the result of the induced muon flux  is $\lsim
10^{-20}\ {\rm cm^{-2}sec^{-1}}$, which is few order of magnitudes
smaller than the upper bound of Super-Kamiokande. This result is
expected since the thermally averaged cross cross is much smaller
than the typical value required by indirect detections.

\section{ Conclusions }
In this paper we considered the supersymmetric $B-L$ model with
inverse seesaw mechanism. We demonstrated that the lightest
right-handed sneutrino in this model can be stable and a viable
candidate for cold dark matter. We studied the relic abundance of
the $B-L$ right-handed sneutrino and showed that the WMAP result,
$\Omega h^2 \simeq 0.11$, can be satisfied in a wide range of the
parameter space. We emphasized that the dominate annihilation
channel of $B-L$ right-handed sneutrino is given by $\tilde{\nu}_1
\tilde{\nu}_1 \to h^0 h^0$, where $h^0$ is the SM-like Higgs. We
also studied the direct detection rate of $B-L$ right-handed
sneutrino. We found that its elastic cross section is consistent
with the upper bounds of current experiments, such as CDMS II and
XENON. Our result of $B-L$ right-handed sneutrino direct detection
is promising and indicates that it can be detectable in near
future experiment.

In addition, we have analyzed the indirect detection rate of $B-L$
right-handed sneutrino. In particular, we focused on the
observation of the Space Observatory PAMELA for positron flux and
also on the neutrino-induced upward through-going muons in the
Super-Kamiokande detector. We showed that although the $B-L$
right-handed sneutrino annihilates at tree level into leptons, the
corresponding cross section is much smaller than the required
one for accommodating PAMELA results. Also the neutrino flux
induced by the $B-L$ right-handed sneutrino annihilations in the
galactic center is much smaller than the Super-Kamiokande's
limits.

Finally it is worth mentioning that our $B-L$ right-handed
sneutrino may be produced at the Large Hadron Collider (LHC)
through the channel is $q\bar q\rightarrow Z_{B-L}\rightarrow
\tilde \nu^{\dagger}_1\tilde \nu_1$. However, the amplitude of
this channel vanishes identically due to the fact that left and
right quarks or leptons has the same $B-L$ quantum numbers.
However, slepton/left-handed sneutrino may decay to right-handed sneutrino, which escapes the detector and gives the missing energy signal similar to other examples of cold DM \cite{progress}.

\section*{Acknowledgments}

We would like to thank Y. Kajiyama for fruitful discussions. S. K.
and H. O. acknowledge partial support from the Science and
Technology Development Fund (STDF) project ID 437 and the ICTP
project ID 30.


\begin{thebibliography}{99}

\bibitem{Spergel:2006hy}
  D.~N.~Spergel {\it et al.}  [WMAP Collaboration],
  Astrophys.\ J.\ Suppl.\  {\bf 170}, 377 (2007).


\bibitem{Ahmed:2009zw}
  Z.~Ahmed {\it et al.}  [The CDMS-II Collaboration],
  Science {\bf 327}, 1619 (2010)
  [arXiv:0912.3592 [astro-ph.CO]].

\bibitem{Khalil:2006yi}
  S.~Khalil,
  J.\ Phys.\ G {\bf 35}, 055001 (2008)
  [arXiv:hep-ph/0611205].

\bibitem{Abbas:2007ag}
  M.~Abbas and S.~Khalil,
  JHEP {\bf 0804}, 056 (2008)
  [arXiv:0707.0841 [hep-ph]].
\bibitem{Emam:2007dy}
  W.~Emam and S.~Khalil,
  Eur.\ Phys.\ J.\  C {\bf 522}, 625 (2007)
  [arXiv:0704.1395 [hep-ph]].
  
  
  \bibitem{b-l-origin}
  R.~N.~Mohapatra and R.~E.~Marshak,
  Phys.\ Rev.\ Lett.\  {\bf 44}, 1316 (1980)
  [Erratum-ibid.\  {\bf 44}, 1643 (1980)];
  F.~Borzumati and A.~Masiero,
  Phys.\ Rev.\ Lett.\  {\bf 57}, 961 (1986);
  C.~S.~Aulakh, A.~Melfo, A.~Rasin and G.~Senjanovic,
  Phys.\ Lett.\  B {\bf 459}, 557 (1999)
  [arXiv:hep-ph/9902409].
  
  
  
  
  
\bibitem{Khalil:2007dr}
  S.~Khalil and A.~Masiero,
  Phys.\ Lett.\  B {\bf 665}, 374 (2008)
  [arXiv:0710.3525 [hep-ph]].

\bibitem{otherb-l}
  K.~Huitu, S.~Khalil, H.~Okada and S.~K.~Rai,
  Phys.\ Rev.\ Lett.\  {\bf 101}, 181802 (2008)
  [arXiv:0803.2799 [hep-ph]],
   L.~Basso, A.~Belyaev, S.~Moretti and C.~H.~Shepherd-Themistocleous,
  Phys.\ Rev.\  D {\bf 80}, 055030 (2009)
  [arXiv:0812.4313 [hep-ph]],
    L.~Basso, A.~Belyaev, S.~Moretti and G.~M.~Pruna,
  JHEP {\bf 0910}, 006 (2009)
  [arXiv:0903.4777 [hep-ph]],
  J.~A.~Aguilar-Saavedra,
  Nucl.\ Phys.\  B {\bf 828}, 289 (2010)
  [arXiv:0905.2221 [hep-ph]],
 P.~Fileviez Perez, T.~Han and T.~Li,
  Phys.\ Rev.\  D {\bf 80}, 073015 (2009)
  [arXiv:0907.4186 [hep-ph]],
 S.~Iso, N.~Okada and Y.~Orikasa,
  Phys.\ Rev.\  D {\bf 80}, 115007 (2009)
  [arXiv:0909.0128 [hep-ph]],
  F.~M.~L.~de Almeida, Y.~A.~Coutinho, J.~A.~Martins Simoes, A.~J.~Ramalho, L.~Ribeiro Pinto, S.~Wulck and M.~A.~B.~do Vale,
  Phys.\ Rev.\  D {\bf 81}, 053005 (2010)
  [arXiv:1001.2162 [hep-ph]],
 P.~Nath {\it et al.},
  Nucl.\ Phys.\ Proc.\ Suppl.\  {\bf 200-202}, 185 (2010)
  [arXiv:1001.2693 [hep-ph]],
P.~Fileviez Perez and M.~B.~Wise,
  Phys.\ Rev.\  D {\bf 82}, 011901 (2010)
  [Erratum-ibid.\  D {\bf 82}, 079901 (2010)]
  [arXiv:1002.1754 [hep-ph]],
  L.~Basso, A.~Belyaev, S.~Moretti and G.~M.~Pruna,
  Phys.\ Rev.\  D {\bf 81}, 095018 (2010)
  [arXiv:1002.1939 [hep-ph]],
 D.~Cogollo, H.~Diniz and C.~A.~de S.Pires,
  Phys.\ Lett.\  B {\bf 687}, 400 (2010)
  [arXiv:1002.1944 [hep-ph]],
 L.~Basso, A.~Belyaev, S.~Moretti, G.~M.~Pruna and C.~H.~Shepherd-Themistocleous,
  arXiv:1002.3586 [hep-ph],
 S.~K.~Majee and N.~Sahu,
  Phys.\ Rev.\  D {\bf 82}, 053007 (2010)
  [arXiv:1004.0841 [hep-ph]],
A.~Ibarra, E.~Molinaro and S.~T.~Petcov,
  JHEP {\bf 1009}, 108 (2010)
  [arXiv:1007.2378 [hep-ph]],
\bibitem{Cao:2010mp}
  Q.~H.~Cao, S.~Khalil, E.~Ma and H.~Okada,
  arXiv:1009.5415 [hep-ph],
  L.~Basso, S.~Moretti and G.~M.~Pruna,
  arXiv:1011.2612 [hep-ph].




\bibitem{Khalil:2010iu}
 S.~Khalil,
  Phys.\ Rev.\  D {\bf 82}, 077702 (2010),
  arXiv:1004.0013 [hep-ph].

\bibitem{inverse-origin}
R. N. Mohapatra, Phys. Rev. Lett. {\bf 56}, 561 (1986);
R. N. Mohapatra and J. W. F. Valle, Phys. Rev. D {\bf 34}, 1642 (1986).

\bi{Khalil:2008ps}
  S.~Khalil and H.~Okada,
  Phys.\ Rev.\  D {\bf 79}, 083510 (2009)
  [arXiv:0810.4573 [hep-ph]].


\bibitem{ArkaniHamed:2000kj}
  N.~Arkani-Hamed, L.~J.~Hall, H.~Murayama, D.~Tucker-Smith and N.~Weiner,
  arXiv:hep-ph/0007001,
  N.~Arkani-Hamed, L.~J.~Hall, H.~Murayama, D.~Tucker-Smith and N.~Weiner,
  Phys.\ Rev.\  D {\bf 64}, 115011 (2001)
  [arXiv:hep-ph/0006312],
  Y.~Grossman and H.~E.~Haber,
  Phys.\ Rev.\ Lett.\  {\bf 78}, 3438 (1997)
  [arXiv:hep-ph/9702421],
  D.~Tucker-Smith and N.~Weiner,
  Phys.\ Rev.\  D {\bf 64}, 043502 (2001)
  [arXiv:hep-ph/0101138],
  D.~Tucker-Smith and N.~Weiner,
  Nucl.\ Phys.\ Proc.\ Suppl.\  {\bf 124}, 197 (2003)
  [arXiv:astro-ph/0208403],
  D.~Tucker-Smith and N.~Weiner,
  Phys.\ Rev.\  D {\bf 72}, 063509 (2005)
  [arXiv:hep-ph/0402065].



\bibitem{Hall:1997ah}
  L.~J.~Hall, T.~Moroi and H.~Murayama,
  Phys.\ Lett.\  B {\bf 424}, 305 (1998)
  [arXiv:hep-ph/9712515],
  L.~E.~Ibanez,
  Phys.\ Lett.\  B {\bf 137}, 160 (1984),
  J.~R.~Ellis, J.~S.~Hagelin, D.~V.~Nanopoulos, K.~A.~Olive and
	M.~Srednicki,
  Nucl.\ Phys.\  B {\bf 238}, 453 (1984),
  J.~S.~Hagelin, G.~L.~Kane and S.~Raby,
	Partners,''
  Nucl.\ Phys.\  B {\bf 241}, 638 (1984),
  K.~Freese,
  Phys.\ Lett.\  B {\bf 167}, 295 (1986).




\bibitem{bl-dm}
 As recent papers of (non-)SUSY $B-L$ DM models, see, {\rm e.g}.,
 S.~Khalil and O.~Seto,
  JCAP {\bf 0810}, 024 (2008)
  [arXiv:0804.0336 [hep-ph]],
   N.~Okada and O.~Seto,
  Phys.\ Rev.\  D {\bf 82}, 023507 (2010)
  [arXiv:1002.2525 [hep-ph]].
T.~Li and W.~Chao,
  Nucl.\ Phys.\  B {\bf 843}, 396 (2011)
  [arXiv:1004.0296 [hep-ph]],
  S.~Kanemura, O.~Seto and T.~Shimomura,
  arXiv:1101.5713 [hep-ph].


\bibitem{Ma:2009gu}
  E.~Ma,
  Phys.\ Rev.\  D {\bf 80}, 013013 (2009)
  [arXiv:0904.4450 [hep-ph]].


\bibitem{Allahverdi:2009ae}
  R.~Allahverdi, B.~Dutta, K.~Richardson-McDaniel and Y.~Santoso,
  Phys.\ Lett.\  B {\bf 677}, 172 (2009)
  [arXiv:0902.3463 [hep-ph]].

\bi{m.carena} M. Carena, A. Daleo, B. A. Dobrescu and T. M. P.
Tait, Phys. Rev. D {\bf 70}, 093009(2004)

\bibitem{Kajiyama:2009ae}
  Y.~Kajiyama, S.~Khalil and M.~Raidal,
  Nucl.\ Phys.\  B {\bf 820}, 75 (2009)
  [arXiv:0902.4405 [hep-ph]].

\bibitem{Bazzocchi:2009kc}
  F.~Bazzocchi, D.~G.~Cerdeno, C.~Munoz and J.~W.~F.~Valle,
  Phys.\ Rev.\  D {\bf 81}, 051701 (2010)
  [arXiv:0907.1262 [hep-ph]].

\bibitem{Dev:2009aw}
  P.~S.~B.~Dev and R.~N.~Mohapatra,
  Phys.\ Rev.\  D {\bf 81}, 013001 (2010)
  [arXiv:0910.3924 [hep-ph]].

\bibitem{griest1}
  K.~Griest,
  Phys.\ Rev.\  D {\bf 38}, 2357 (1988)
  [Erratum-ibid.\  D {\bf 39}, 3802 (1989)]
  [Phys.\ Rev.\  D {\bf 39}, 3802 (1989)].

\bibitem{Griest:1989zh}
  K.~Griest, M.~Kamionkowski and M.~S.~Turner,
  Phys.\ Rev.\  D {\bf 41}, 3565 (1990).


\bibitem{Antoniadis:2010nb}
  I.~Antoniadis, E.~Dudas, D.~M.~Ghilencea and P.~Tziveloglou,
  arXiv:1012.5310 [hep-ph].


\bibitem{Aprile:2010um}
  E.~Aprile {\it et al.}  [XENON100 Collaboration],
  Phys.\ Rev.\ Lett.\  {\bf 105}, 131302 (2010)
  [arXiv:1005.0380 [astro-ph.CO]].
\bibitem{Aprile:2011ts}
  E.~Aprile {\it et al.}  [XENON100 Collaboration],
  arXiv:1104.3121 [astro-ph.CO].



\bibitem{deCarlos:1997yv}
  B.~de Carlos and J.~R.~Espinosa,
  Phys.\ Lett.\  B {\bf 407}, 12 (1997)
  [arXiv:hep-ph/9705315].


\bi{langacker}
 V. Barger, P. Langacker, lan Lewis, Mat McCaskey, G. Shaughnessy and B. Yencho, Phys.Rev. {\bf D75}: 115002,2007.


\bi{gondolo}
P. Gondolo, J. Edsjo, P. Ullio, L. Bergstrom, Mia Schelke and E.A. Baltz, JCAP 0407: {\bf008}, 2004.

\bibitem{Goodman:1984dc}
  M.~W.~Goodman and E.~Witten,
  Phys.\ Rev.\  D {\bf 31}, 3059 (1985).

\bibitem{Arina:2007tm}
  C.~Arina and N.~Fornengo,
  JHEP {\bf 0711}, 029 (2007)
  [arXiv:0709.4477 [hep-ph]].


\bibitem{Bernabei:2008yi}
  R.~Bernabei {\it et al.}  [DAMA Collaboration],
  Eur.\ Phys.\ J.\  C {\bf 56}, 333 (2008).

\bibitem{Khalil:2010yt}
  S.~Khalil, H.~S.~Lee and E.~Ma,
  Phys.\ Rev.\  D {\bf 81}, 051702 (2010)
  [arXiv:1002.0692 [hep-ph]];
 H.~S.~Lee,
  AIP Conf.\ Proc.\  {\bf 1078}, 569 (2009)
  [arXiv:0808.3600 [hep-ph]].



\bibitem{Adriani:2008zr}
O.~Adriani {\it et al.},
Nature {\bf 458}  (2009) 607,
O.~Adriani {\it et al.},
  Phys.\ Rev.\ Lett.\ {\bf 102} (2009) 051101.


\bibitem{Abdo:2009zk}
A.A.~Abdo {\it et al.},
Phys.\ Rev.\ Lett.\ {\bf 102} (2009) 181101,
M.~Ackermann {\it et al.},
arXiv:1008.3999 [astro-ph.HE].


\bibitem{bw}
  M.~Ibe, H.~Murayama and T.~T.~Yanagida,
  Phys.\ Rev.\  D {\bf 79}, 095009 (2009)
  [arXiv:0812.0072 [hep-ph]],
   D.~Feldman, Z.~Liu and P.~Nath,
  Phys.\ Rev.\  D {\bf 79}, 063509 (2009)
  [arXiv:0810.5762 [hep-ph]].

\bibitem{sf}
J.~Hisano, S.~Matsumoto, M.~M.~Nojiri and O.~Saito,
  Phys.\ Rev.\  D {\bf 71}, 063528 (2005)
  [arXiv:hep-ph/0412403],
  M.~Pospelov and A.~Ritz,
  Phys.\ Lett.\  B {\bf 671}, 391 (2009)
  [arXiv:0810.1502 [hep-ph]],
J.~L.~Feng, M.~Kaplinghat and H.~B.~Yu,
  Phys.\ Rev.\  D {\bf 82}, 083525 (2010)
  [arXiv:1005.4678 [hep-ph]].




\bibitem{Hisano:2008ah}
  J.~Hisano, M.~Kawasaki, K.~Kohri and K.~Nakayama,
  Phys.\ Rev.\  D {\bf 79}, 043516 (2009)
  [arXiv:0812.0219 [hep-ph]].


\bibitem{Beacom:2006tt}
  J.~F.~Beacom, N.~F.~Bell and G.~D.~Mack,
  Phys.\ Rev.\ Lett.\  {\bf 99}, 231301 (2007)
  [arXiv:astro-ph/0608090];
  H.~Yuksel, S.~Horiuchi, J.~F.~Beacom and S.~Ando,
  Phys.\ Rev.\  D {\bf 76}, 123506 (2007)
  [arXiv:0707.0196 [astro-ph]].


\bibitem{Jungman:1995df}
As a review, see  G.~Jungman, M.~Kamionkowski and K.~Griest,
  Phys.\ Rept.\  {\bf 267}, 195 (1996).

\bibitem{Ritz:1987mh}
  S.~Ritz and D.~Seckel,
  Nucl.\ Phys.\  B {\bf 304}, 877 (1988).

\bibitem{Navarro:1995iw}
  J.~F.~Navarro, C.~S.~Frenk and S.~D.~M.~White,
  Astrophys.\ J.\  {\bf 462}, 563 (1996)
  [arXiv:astro-ph/9508025].


\bibitem{Desai:2004pq}
  S.~Desai {\it et al.}  [Super-Kamiokande Collaboration],
  Phys.\ Rev.\  D {\bf 70}, 083523 (2004)
  [Erratum-ibid.\  D {\bf 70}, 109901 (2004)]
  [arXiv:hep-ex/0404025].




\bibitem{Goodman:1984dc}
  M.~W.~Goodman and E.~Witten,
  Phys.\ Rev.\  D {\bf 31}, 3059 (1985).

\bibitem{Arina:2007tm}
  C.~Arina and N.~Fornengo,
  JHEP {\bf 0711}, 029 (2007)
  [arXiv:0709.4477 [hep-ph]].

\bibitem{progress}
In preparation.

\end{thebibliography}
\end{document}